\documentstyle[aps,preprint]{revtex}
\input{epsf}

\begin{document}
\title{Kinetics of Ordering in Fluctuation-Driven First-Order Transitions: Simulations
and Dynamical Renormalization}
\author{N. A. Gross, College of General Studies at Boston University}
\author{M. Ignatiev and Bulbul Chakraborty, Physics Department at Brandeis University}
\date{\today}
\maketitle
\begin{abstract}
Many systems where interactions compete with each other or with
constraints are well described by a model first introduced by
Brazovskii. Such systems include block copolymers, alloys with
modulated phases, Rayleigh-Benard Cells and type-I
superconductors. The hallmark of this model is that the fluctuation
spectrum is isotropic and has a minimum at a nonzero wave vector
represented by the surface of a d-dimensional hyper-sphere. It was
shown by Brazovskii that the fluctuations change the free energy
structure from a $ \phi ^{4}$ to a $\phi ^{6}$ form with the
disordered state metastable for all quench depths. The transition from
the disordered to the periodic, lamellar structure changes from second
order to first order and suggests that the dynamics is governed by
nucleation. Using numerical simulations we have confirmed that the
equilibrium free energy function is indeed of a $ \phi ^{6}$ form. A
study of the dynamics, however, shows that, following a deep quench,
the dynamics is described by unstable growth rather than nucleation. A
dynamical calculation, based on a generalization of the Brazovskii
calculations shows that the disordered state can remain unstable for a
long time following the quench.
\end{abstract}
\pacs{05.10.Cc,05.40.-a,05.45.-a}
\section{Introduction}

Kinetics of growth can be broadly classified into two categories; nucleation
and spinodal decomposition (or continuous ordering). The former applies to
situations where the initial state is metastable and the latter to where the
initial state is unstable. The distinction becomes unclear near the limit of
metastability and in systems where the concept of metastability itself is
ill-defined. A class of systems where the definition of metastability
becomes ambiguous is one where there is a fluctuation-driven first-order
phase transition. A notable example of a system undergoing such a
transition is a symmetric di-block copolymer. These systems undergo
microphase separation and there is a temperature-driven transition from a
uniform phase to a lamellar phase\cite{Bib:Blockcopolymer}. Another system
which is governed by the same dynamical equation as block copolymers, and so
exhibits a similar transition, is a Rayleigh-Benard Cell \cite{Bib:RBreview}.

A theoretical model describing these transitions was proposed by Brazovskii
in 1975\cite{Braz1975}. Di-block copolymers were shown to belong to the
Brazovskii class by Leibler\cite{Bib: Lieb80} and the theoretical
predictions regarding the equilibrium nature of the ordering transition have
been experimentally verified\cite{Bib:coployexpt}. It was shown by
Brazovskii, within a self-consistent Hartree approximation, that the
fluctuations destroy the mean-field instability and lead to a first-order
phase transition\cite{Braz1975,FB1989}. Theories of nucleation and growth
have been constructed based on the idea that the static,
Brazovskii-renormalized theory can provide an effective potential for a
stochastic Langevin equation\cite{FB1989,HS1995}. In this paper we examine
the validity of this description by using numerical simulations to study the
relaxational dynamics of a model described by the Brazovskii Hamiltonian and
comparing the results to the predictions of the ``static'' nucleation
theories and to the predictions of a dynamically renormalized theory\cite
{ICG99}.

\section{The Equation of Motion and Numerical Simulations}

Our starting point is the same as in Swift and Hohenberg\cite{HS1995} and
describes a system with model A dynamics\cite{HHreview} and a Brazovskii
Hamiltonian such as would apply to the diblock copolymers\cite{FB1989}. The
Brazovskii Hamiltonian is characterized by a fluctuation spectrum whose
maximum occurs at a non-zero wave vector $|{\bf q}|=q_{0}$ and can be
represented by a hypersphere in $d-$ dimensions. The full form of the
Hamiltonian is 
\begin{equation}
H=\int d{\bf {x}}[\frac{1}{2}\phi (\nabla ^{2}+q_{0}^{2})^{2}\phi +\frac{1}{2}
\tau \phi ^{2}+\frac{1}{4!}\lambda \phi ^{4}]  \label{Free Energy}
\end{equation}
\newline
The dynamics is taken to be relaxational and so the equation of motion is
given by the Langevin Equation. 
\begin{equation}
\frac{d\phi }{dt}=-M\frac{\delta H}{\delta \phi }+\eta  \label{Lang}
\end{equation}
\newline
where $M$ is the mobility (which sets the time scale for the problem) and $
\eta $ is Gaussian noise ($<\eta >=0,$ $<\eta (\vec{x},t)$ $\eta (\vec{x}
^{\prime },t^{\prime })>=\delta (\vec{x}-\vec{x}^{\prime })\delta
(t-t^{\prime })$ ).

The stochastic Langevin equation derived from the Hamiltonian differs
from the usual Ginzburg-Landau description\cite{HHreview} because of
the appearance of the unusual gradient term. This dynamical equation
is usually referred to as the Swift-Hohenberg equation and falls into
the Type $I_s$ classification of Cross and Hohenberg
\cite{bib:CHreview}. In this classification the system is unstable
to a static, spatially periodic structure.

The complete equation of motion is; 
\begin{equation}
\frac{d\phi }{dt}=-M(q_{0}^{2}\nabla ^{2}\phi +\nabla ^{4}\phi
+(q_{0}^{4}+\tau )\phi +\frac{\lambda }{6}\phi ^{3})+\eta
\label{Equation of Motion}
\end{equation}
\newline
In Eq. (\ref{Equation of Motion}), the coupling constant, $\lambda$ has been
rescaled by the noise strength. For systems where the noise strength is
small, such as Rayleigh-Benard convection\cite{Hoh92}, the effective
coupling constant is also small. In di-block copolymers, the noise strength
is of the order of $k_B T$.

For numerical calculations, the Langevin equation must be approximated by a
discrete equation. 
\begin{equation}
\phi (t+\Delta t)=\phi (t)-M\Delta t(\pounds \phi +\pounds ^{2}\phi
+(r+q_{0}^{4})\phi +\frac{\lambda }{6}\phi ^{3})+(M\Delta t/(\Delta
x)^{3})^{1/2}\eta  \label{Discrete EOM}
\end{equation}
\pounds\ here is the discrete Laplacian. This can take on several forms
though for this work we use the simplest form: $\frac{d^{2}\phi }{dx^{2}}
\approx \pounds \phi =\sum \frac{1}{\Delta x^{2}}(x+\Delta x)+\phi (x-\Delta
x)-2\phi (x))$ where the sum is over the dimensions of the lattice. Other
choices which include next nearest or more complicated neighborhoods are
possible\cite{OP1987}. This is important if isotropy is of concern however
the effect is small and can be ignored. The scaling of the noise in Eq. \ref
{Discrete EOM} reflects the fact that as the cell size or time steps become
smaller (larger), the possible fluctuations become larger (smaller).

For these simulations we set $q_{0}=1$ and choose the lattice spacing such
that one lamellar spacing spans six lattice sites ($\Delta x=6/2\pi $) For
most of this work the overall lattice size is $60^{3}$. The mobility, $M$,
is set to one. In order for the simulations to be numerically stable the
time scale must be smaller then some stability time, $\Delta t\sim 1/\Delta
x^{4}.$ To satisfy this the timescale is chosen to be $\Delta t$ = 0.01 and
measurements are taken at intervals of $100*\Delta t$ or longer.

Previous theories of nucleation have analyzed the above equation under
the approximation that the effects of fluctuations can be incorporated
into a {\it static} renormalized free energy $F_{R}$ which replaces
the bare free energy $F$ in Eq. (\ref{Lang}) \cite{FB1989,HS1995}. We
will compare our numerical results to the Brazovskii predictions for
the static renormalized parameters and show that the static scenario
is beautifully borne out by the simulations in three dimensions. We
will then analyze the time-dependence of the structure factor as
observed in the numerical simulations and show that this time
dependence is inconsistent with the dynamical picture based on a {\it
static} renormalized free energy. A theory based on a {\it dynamical}
renormalization of Eq. (\ref{Lang}) can provide a qualitatively
correct description of the numerical simulations.

\section{Statics}

\subsection{Brazovskii Theory}

\begin{figure}[h]
\epsfxsize=7in \epsfysize=1in
\epsfbox{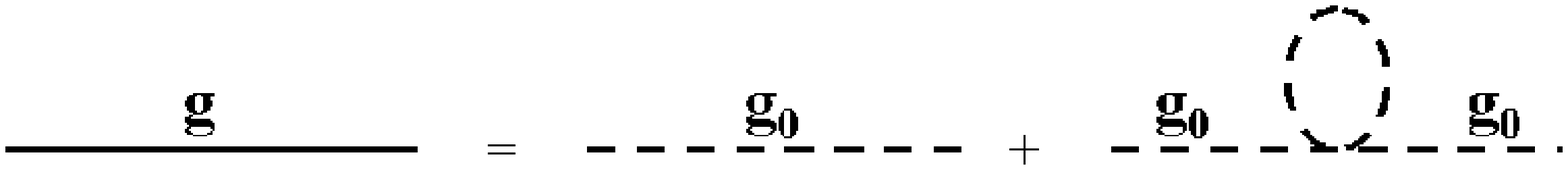}
\vspace{0.3in}
\caption{Diagrams used for static
renormalization of the mass parameter in the free energy. The Hartree
approximation only uses diagrams up to order $\lambda $ and then replaces the
bare parameter in the integrand with the renormalized one to solve the
equations self consistently.
}
\label{Fig: Static One Loop}
\end{figure}

Brazovski's treatment of the model within the Hartree approximation can be
restated in terms of an expansion of the thermodynamic potential $\Gamma ( 
\bar{\phi})$, the generating functional for the vertex functions\cite
{Amit,Zinn-Justin}. This approach has been described in detail by
Fredrickson and Binder\cite{FB1989}, and within the Hartree approximation
leads to a renormalized mass term ($r$) in Eq.(\ref{Free Energy}). The
diagrams up to one loop are shown in Fig. \ref{Fig: Static One Loop}, and
the mass renormalization relation is given by; 
\begin{equation}
r+(q^{2}-q_{0}^{2})^{2}=\tau +(q^{2}-q_{0}^{2})^{2}+{{\frac{\lambda }{2}}}
\int {{\frac{d{\bf q}}{(2\pi )^{d}}}}(\tau +(q^{2}-q_{0}^{2})^{2})^{-1}
\label{Eq: Renorm1}
\end{equation}

This approximation is made self consistent by replacing the bare parameter
in the integrand by the renormalized parameter. Essentially the bare
propagator on the loop in Fig. (\ref{Fig: Static One Loop}) is replaced by
renormalized propagator. The leads to the Hartree result.

\begin{equation}
r=\tau +{{\frac{\lambda }{2}}}\int {{\frac{d{\bf q}}{(2\pi )^{d}}}}
(r+(q^{2}-q_{0}^{2})^{2})^{-1}=\tau +\alpha \lambda /\sqrt{r}  \label{rbar1}
\end{equation}
\newline
where $\alpha $ is proportional to the surface area of a $d$ dimensional
sphere of unit radius. According to Brazovski\cite{Braz1975} this
approximation is good only for $\lambda ^{-6}<<1.$

The interesting point about Eq. (\ref{rbar1}) is that even for negative
values of the bare parameter, $\tau $, the renormalized parameter, $r$, is
positive. This implies that the disordered phase is always metastable.
Brazovskii went on to show that the bare coupling parameter $\lambda$ gets
renormalized to a {\it negative} value leading to a $\phi^6$ theory and the
possibility of a first-order phase transition\cite{FB1989}.

\subsection{Computational Results.}

The numerical simulations can measure the static structure factor in the
disordered phase which is predicted by the Brazovskii theory \cite
{Braz1975,FB1989} to be, 
\begin{equation}
S(q)^{-1}=r+(q^{2}-q_{0}^{2})^{2}  \label{r_bar}
\end{equation}
where $r$ is the renormalized control parameter \cite{Braz1975,HS1995}.
Thus, $S(q_{0})^{-1}$ is just $r$. The renormalized mass can, therefore, be
measured by monitoring the peak of the structure factor. The Hartree
calculations predict $r$ to be positive and asymptotically approaching zero
as $\tau \rightarrow -\infty $. Experiments in symmetric di-block copolymers
have verified that the behavior of $S(q_{0})$ is consistent with the
Brazovskii predictions and is very different from the mean-field prediction
( $S(q_{0})=\tau ^{-1}$).

\begin{figure}[h]
\epsfxsize=5.1in \epsfysize=3.2in
\epsfbox{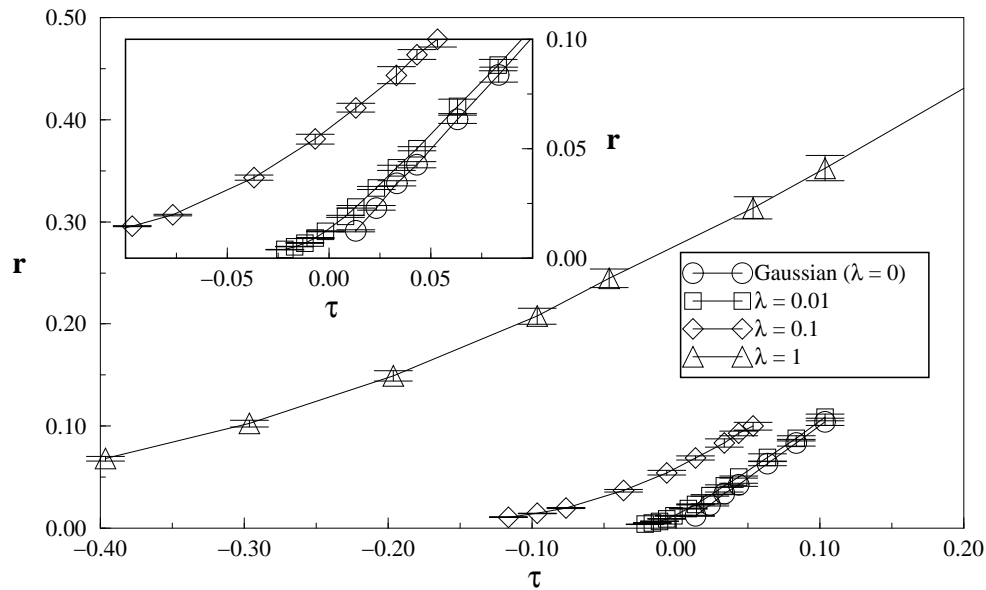}
\vspace{0.3in}
\caption{The renormalized control
parameter, $r$, against the bare control parameter, $\tau $. The different
data sets are for different values of $\lambda $. The Gaussian case, $
\lambda =0,$ which was used to set the scale, is also plotted for
comparison. The inset shows the region near the mean field transition. Note
that the renormalized parameter does not cross zero here. }
\label{Fig: r_vs_t}
\end{figure}

Fig. \ref{Fig: r_vs_t} shows results for $r$ obtained from our simulations
for different values of the coupling constant, $\lambda $. The Gaussian case
($\lambda =0$) is also shown for comparison. The system was run for 1000
time steps (each time step being $100*\Delta t$) and 100 samples were taken.
The data points plotted are the average of the samples while the error bars
represent the standard deviation. For large positive values of the bare
control parameter, $\tau $, the fourth order term becomes less important and 
$r$ approaches the Gaussian value. This is seen clearly for small values of $
\lambda $. As $\tau $ approaches zero, the measured values of $r$ deviate
from the Gaussian case and stay positive for all $\tau $. Some care was
taken to normalize the values presented in Fig. \ref{Fig: r_vs_t}. For the
Gaussian case, $S(q_{0})^{-1}$ should be linearly related to $\tau $ and the
slope of the line should be one. In calculating $S(q_{0})$ several
normalizations are needed including the normalization due to the Fourier
Transform (FT) and the normalization due to circular averaging. While the FT
normalization is just related to the system size, the normalization due to
circular averaging is more complicated to calculate. Instead of calculating
these normalizations directly, the slope of the raw result from the Gaussian
case was used to normalize all of the data. Another concern is that for an
infinite continuous system this line should have an x-intercept at zero;
however, for the finite systems on a grid used in the simulations the
x-intercept is slightly negative. To account for this, all of the data is
shifted over by an amount equal to that intercept. The importance of this
shift will become apparent in the next paragraph when a scaling is applied
and the values of the results around zero are magnified. A small negative
(positive) value becomes a much larger negative (positive) value and so
shifts from negative to positive will become important.

In reference\cite{HS1995} the authors show that, within the Hartree
approximation, the $r(\tau )$curves for different $\lambda ^{\prime }s$
could be described by a single functional form in terms of scaled variables $
\tau ^{*}$ and $r^{*}$; 
\begin{eqnarray}
r^{*} &=&r*(\alpha \lambda )^{-2/3}  \label{r Scaling} \\
\tau ^{*} &=&\tau *(\alpha \lambda )^{-2/3}  \nonumber
\end{eqnarray}

\begin{figure}[h]
\epsfxsize=5.1in \epsfysize=3.2in
\epsfbox{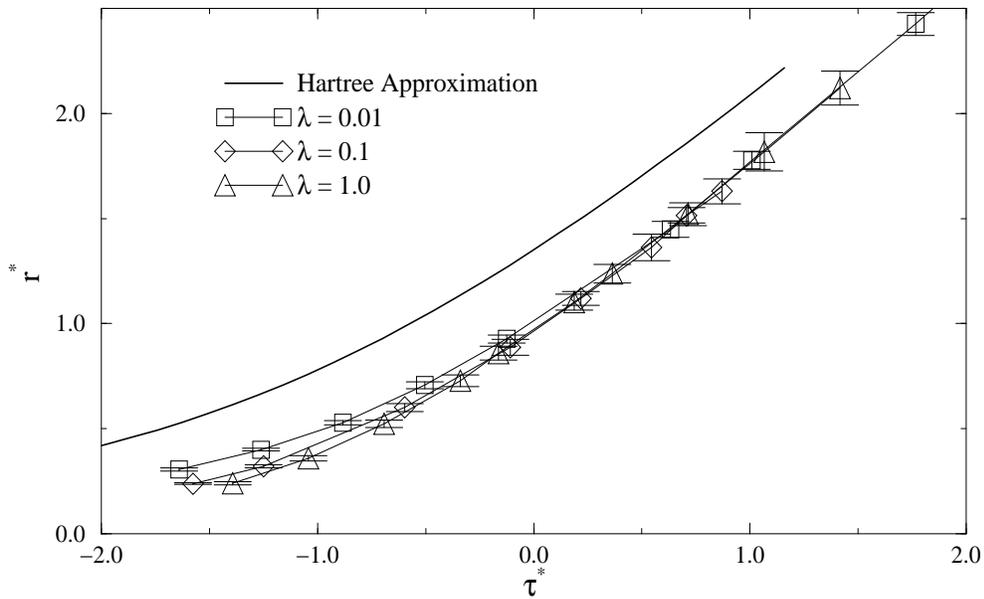}
\vspace{0.3in}
\caption{Scaling of $r(\tau )$ using the
form deduced from Hartree theory (\textit{cf} text). The data collapses well
except at the deepest quenches. The scaling curve predicted by the Hartree
approximation is shown for comparison. It is in fairly good agreement with
the data considering that the values of $\lambda $ used in the simulations
are four order of magnitude larger then that for which the the
approximation is expected to be valid.}
\label{Fig: r* vs t}
\end{figure}

Figure \ref{Fig: r_vs_t} shows plots of $r^{*}$ against $\tau
^{*}$obtained from simulations for the three different values of
$\lambda$. The curves are seen to scale quite well but the scaled
curve falls significantly below the theoretical prediction
\cite{Braz1975,FB1989,HS1995} shown as the thick black line. The
theoretical prediction that the value of $r^{*}$(and $r$) never
becomes negative and is asymptotic to zero is, however, clearly borne
out by the simulations. The deviation from the theoretical line could
be a system size effect, however, results for larger and smaller
systems are consistent with the data shown in Fig.\ref{Fig: r* vs
t}. For the Gaussian case $r^{*}=\tau ^{*}$ and for large positive
values of the bare parameter both the theory and simulations approach
this. Since the theory is larger then the simulations and both are
above the Gaussian, our simulations suggest that the contributions to
the two point function from diagrams not included in the Hartree
approximation serve to lower the overall correction to mean field
theory. Another property that the simulation data exhibits in
Fig. \ref{Fig: r* vs t} which is not predicted by the theory is that
the curves for different values of $\lambda$ diverge from each other
at negative values of $\tau $, {\it i.e.} the scaling is not
perfect. Smaller values of $ \lambda $ lie closer to the theory as
expected. The computations were done using values of $\lambda \sim
10^{-2}$ and larger while the theory is valid only for $\lambda \sim
10^{-6}$ so the theory's scaling predictions seem to be far more
robust then expected.

\begin{figure}[h]
\epsfxsize=5.1in \epsfysize=3.2in
\epsfbox{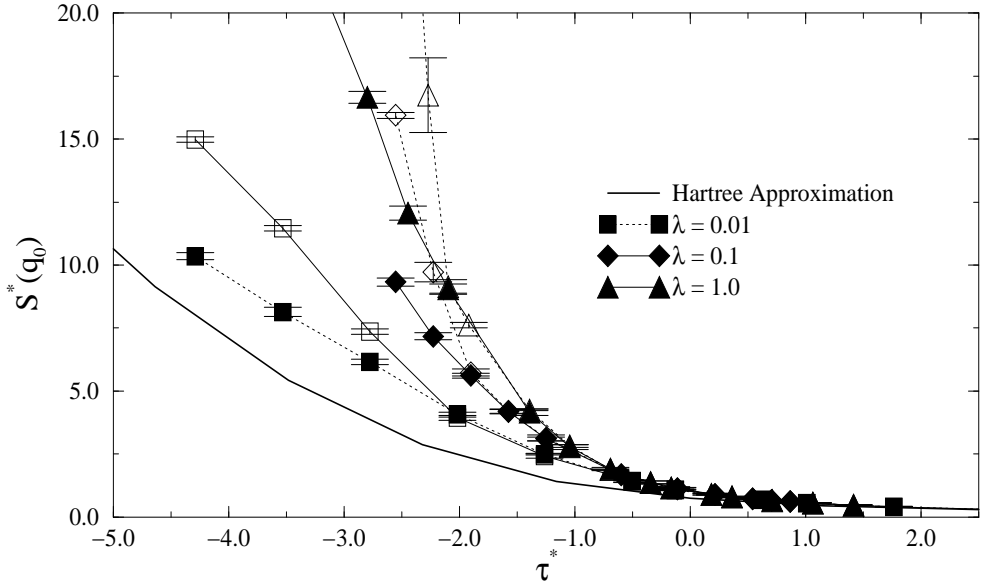}
\vspace{0.3in}
\caption{ Temperature dependence of the 
average intensity at  the peak of the structure factor.
To show the transition, results from 
two different starting configurations are compared.
The shapes of the symbols denote different values of $\lambda $, whereas the
shading distinguishes between the starting configurations.
The closed
symbols represent systems that were started in a disordered state,
simulating a hot start, while the open symbols are for systems which were
started with an ordered lamellar structure already present. At high
temperatures the two starting conditions give nearly the same results while
at low temperatures the systems prepared in the ordered state have higher
peak intensities implying that the systems prepared in the disordered states
are in a metastable state or not in equilibrium ({\it cf} text).
}
\label{Fig: Spinodal}
\end{figure}

To further explore the nature of the phase transition in this model, we can
compare scaled values of $S(q_{0})$ obtained from a hot (random) and cold
(purely modulated order) initial configurations. Fig. \ref{Fig: Spinodal}
shows just such a plot. As in the previous plot, the theory is shown as a
thick black line. For each value of $\lambda $ there are two lines
presented: one for a disordered start and one for an ordered start. The
averaging was done as described above. For positive and small negative
values of $\tau $ the results from the disordered start and the ordered
start are nearly the same. For $\tau ^{*}$ below some $\tau _{s}^{*}$ the
two differ with the ordered phase having a higher peak. It should be pointed
out that for these values of $\tau ^{*}$ the disordered state is not in
equilibrium (metastable or stable)and there is a slow but definite time
evolution of the structure factor over the time period in which the averages
are taken. This time-evolution will be discussed in more detail in the
section on dynamics. For now the disordered start points are included in the
plot simply as a comparison to see where the ordered phase melts. The value
where the two data sets diverge can be considered an estimate for the limit
of stability (spinodal) of the lamellar phase. As a consistency check the
average value of the wave vector was measured. For values of $\tau $ above $
\tau _{s}$ the wave vector is small and points in a random direction, while
for lower values of $\tau $ the average wave vector is large and and points
along the direction that the system was prepared in. For all three sets of
data the lamellar spinodal lies in the range of $-2.1<\tau _{s}^{*}<-1.9$.
This is consistent with the prediction from Hartree theory that the
first-order transition to the lamellar phase occurs at a lower value of $
\tau^*$, ${\tau}_{t}^* = -2.74$\cite{HS1995}. That is to say that our
measured value of the lamellar spinodal is above the predicted transition
temperature as it should be. For values of $\tau^*$ below $\tau _{s}^{*}$,
however, systems prepared in the disordered phase always evolve towards the
ordered phase. It is possible that the lamellar ordered phase is enhanced by
the boundary conditions - the lattice size is always chosen to be
commensurate with the lamellar wavelength - however, it should be noted that
lamella form in many possible directions where the system size is not
commensurate with the lamellar wavelength. Discussion in the dynamics
sections should shed some light on this issue.

\section{Dynamics}

\subsection{Relaxational Dynamics Based on Static Renormalized Parameters}

Previous analysis of nucleation and metastability, in the context of
fluctuation-driven phase transitions, have been based on a Langevin equation
with the force obtained from the static, Hartree-renormalized free energy
function\cite{FB1989,HS1995}. Fredrickson and Binder\cite{FB1989} used this
approach to compute the nucleation barrier and the completion rate of
nucleation and growth in di-block copolymers. The interfacial tension was
found to be small leading to a small nucleation barrier and rapid nucleation
for deep quenches. In this picture, there is no essential difference between
the kinetics of a fluctuation-driven first-order transition and a weak-first
order transition. The results of our simulations suggest a very different
scenario for the growth of the lamellar structures in a Brazovskii model.
The shape of the nucleating droplet was more carefully analyzed by Swift and
Hohenberg\cite{HS1995} by taking into account spatial inhomogeneities in the
effective free energy function. Their analysis relied on constructing a
coarse-grained free energy functional. The coarse graining was based on a
momentum-shell renormalization idea where fluctuations with momenta far away
from the shell defined by $|{\bf q| = q_0}$ are successively integrated out.
This analysis led to non-spherical droplets and were consistent with the
picture of spinodal nucleation that had been obtained earlier\cite
{bib:Kleinspinodal}.

As pointed out in the work of Swift and Hohenberg\cite{HS1995}, a complete
theory of the dynamics of fluctuation-driven first order transitions would
have to be based on a coarse-graining of the full dynamics as expressed by
the original Langevin equation with the bare Brazovskii Hamiltonian. In this
paper, we compare the results of our numerical simulations to the
predictions of a Hartree-renormalization of the full dynamical equation as
expressed in Eq. (\ref{Lang}). Our emphasis has been on understanding the
nature of the dynamics of the metastable phase and we have not analyzed, in
any detail, the spatial structures associated with the growth process.

\subsection{Dynamical Renormalization and Simulation Results}

In order to systematically analyze the effects of fluctuations on the
kinetics of growth of the lamellar phase, perturbative techniques analogous
to the static Hartree approximation have to be applied to the Langevin
equation. This is most conveniently done through the dynamical-action
formalism\cite{Zinn-Justin}. The application of this method to the S-H
equation is outlined by Ignatiev et. al. \cite{ICG99}. In the
dynamical-action formalism, the average value of an operator $O(\phi )$ over
the noise history is rewritten as a functional integral; 
\begin{equation}
\langle O\{\phi ({\bf x},t)\}\rangle =\int D{\phi }\exp {(-S[{\phi }])}~,
\label{Generating Functional 1}
\end{equation}
where $S[{\phi }]$ is the dynamical action\cite{Zinn-Justin}.

As in the static case, the calculation of dynamical correlation functions is
most conveniently formulated through the construction of a generating
function\cite{Zinn-Justin,Amit}. To establish the closest analogy with the static
calculations, it is useful to work with the generator of vertex functions, ${
\ \ \ \Gamma }_{dyn}[\bar{\phi}]$, and establish a diagrammatic expansion
which is the exact analog of the diagrams retained by Brazovskii in the
static calculation. The simulation results demonstrate that the static
properties of the S-H equation closely follow the Brazovskii predictions
and, therefore, it seems appropriate to apply this approximation scheme to
the dynamics. The correspondence between statics and dynamics becomes
particularly transparent in a super-field formulation\cite{Kurchan} which
shows that the dynamical perturbation theory in terms of the super fields
has exactly the same structure as the static perturbation theory except for
the appearance of a very different ``kinetic'' term which leads to a {\it 
bare} propagator that is distinct from the static theory. The super-field
correlation functions can be calculated by constructing diagrams as in the
static theory but replacing the static bare propagator by the appropriate
dynamical one\cite{Kurchan}. The super-field correlation functions encode
all the dynamical correlations of the field $\phi $ and the latter can be
extracted from well-defined relations\cite{Kurchan}. The details of this
calculation and the results will be described in a separate publication\cite
{ICG99}. In this paper we present the main results which can be compared to
the numerical simulations.

\begin{figure}[h]
\epsfxsize=7in \epsfysize=1in
\epsfbox{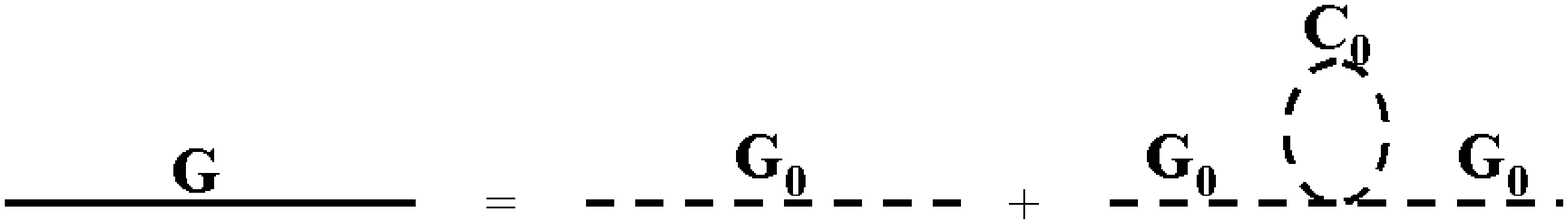}
\vspace{0.3in}
\caption{As with the static case,
diagrams only up to order $\lambda $ are used and then the bare parameter is
replaced by the renormalized parameter to give a self consistent solution.
}
\label{Fig: Dynamic One Loop}
\end{figure}

The two correlation functions which appear in the dynamical study are 
\[
G(t,t^{\prime })=<\phi (t)\eta (t^{\prime })> 
\]
and 
\[
C(t,t^{\prime })=<\phi (t)\phi (t^{\prime })> 
\]
Both of these can be obtained from a single super-field correlation function
($Q(1,2)$) which is related to $\Gamma _{dyn}$ through 
\begin{equation}
Q^{-1}(1,2)={\frac{{\delta }^{2}\Gamma _{dyn}}{\delta \Phi _{1}\delta \Phi
_{2}}}  \label{super}
\end{equation}
where $\Phi $'s are the super fields and the derivative is evaluated with
the source term set to zero. Applying this formalism to Eq. (\ref{Lang}),
leads to an expression for $\Gamma _{dyn}$ which involves two-point
correlation functions of the fluctuations and is a natural extension of the
static $\Gamma (\bar{\phi})$ to correlation functions involving time. The
expression, however, does not lend itself to easy analysis except in two
cases, early times when a variant of linear theory can be applied and late
times when the system is stationary. The late time dynamics involves the
analysis of the nonlinear terms and will be discussed in the concluding
section. The early time dynamics is where one is justified in retaining only
quadratic terms in the renormalized $\Gamma _{dyn}$. In this limit we obtain
the following equation for the equal-time correlation function $
C_{q}(t,t)=<\phi _{q}(t)\phi _{-q}(t)>$ which is the structure factor that
is monitored in the simulations:

\begin{eqnarray}
C_{q}(t,t) &=&{\int }_{0}^{t}[G(t,t^{\prime })]^{2}dt^{\prime }
\label{Eq: Correlation Functions} \\
G_{q}(t,t^{\prime }) &=&\int_{t}^{t^{\prime }}(r(t^{\prime \prime}
)+(q^{2}-q_{0}^{2})^{2})dt^{\prime \prime} ~ . \nonumber
\end{eqnarray}
The
mass parameter, $r(t)$, which is now time-dependent, is renormalized in the dynamic theory using the same
approximation as in the static theory. The diagrams used are given in fig. 
\ref{Fig: Dynamic One Loop}. These again are only to one loop. The mass term
then becomes

\begin{eqnarray}
r(t)+(q^{2}-q_{0}^{2})^{2} &=&\tau +(q^{2}-q_{0}^{2})^{2}+ \\
&&{{\frac{\lambda }{2}}}[\int {{\frac{d{\bf q}}{(2\pi )^{d}}}}\frac{1}{
(\tau +(q^{2}-q_{0}^{2})^{2})}-\int {{\frac{d{\bf q}}{(2\pi )^{d}}}} 
\frac{\exp \{-2(\tau +(q^{2}-q_{0}^{2})^{2})t\}}{(\tau
+(q^{2}-q_{0}^{2})^{2})}]  \nonumber
\end{eqnarray}
Again, the approximation is made self consistent by replacing the
$\tau$ with $r$ in the integrand. 
\begin{equation}
r(t)=r_{eq}-{{\frac{\lambda }{2}}}\int {{\frac{d^{3}q}{(2\pi )^{3}}\frac{
\exp (-2D(q)t)}{D(q)}}}  \label{r(t)}
\end{equation}
$D(q)$ here is the renormalized propagator, $
D(q)=r_{eq}+(q^{2}-q_{0}^{2})^{2}$ where $r_{eq}$ denotes the {\it static}
renormalized mass parameter which is the solution to eq. \ref{r_bar}.

In Eq. \ref{r(t)}, at long times the integrand in the second term becomes
small and $r(t)$ approaches $r_{eq}.$ which is confirmed in the section on
statics above. At time zero the subtraction of the second term is just a
subtraction of the Hartree level correction factor to the bare parameter
introduced earlier Eq. (\ref{rbar1} and, therefore, $r(t=0)$ is just the
bare parameter $\tau$, while as $t \rightarrow \infty$, $r(t)$ approaches
the renormalized, equilibrium value.

\begin{figure}[h]
\epsfxsize=5.1in \epsfysize=3.2in
\epsfbox{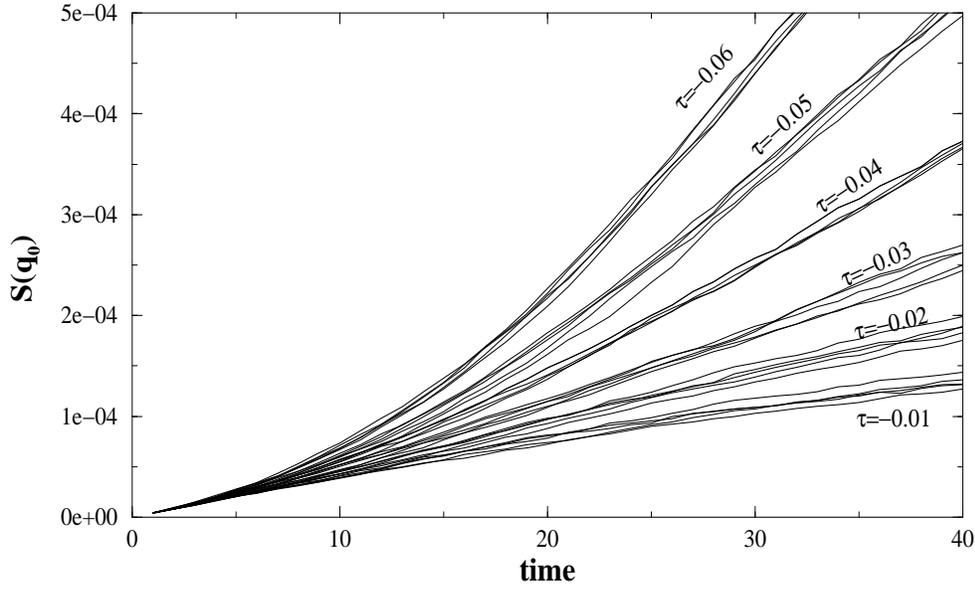}
\vspace{0.3in}
\caption{The time evolution of the structure
factor peak for the  earliest times. Different sets are for different
quench depths as indicated on the graph. All quenches to below the mean
field transition temperature initially show unstable growth.
}
\label{Fig: S(q0) shorttimes}
\end{figure}

\begin{figure}[h]
\epsfxsize=5.1in \epsfysize=3.2in
\epsfbox{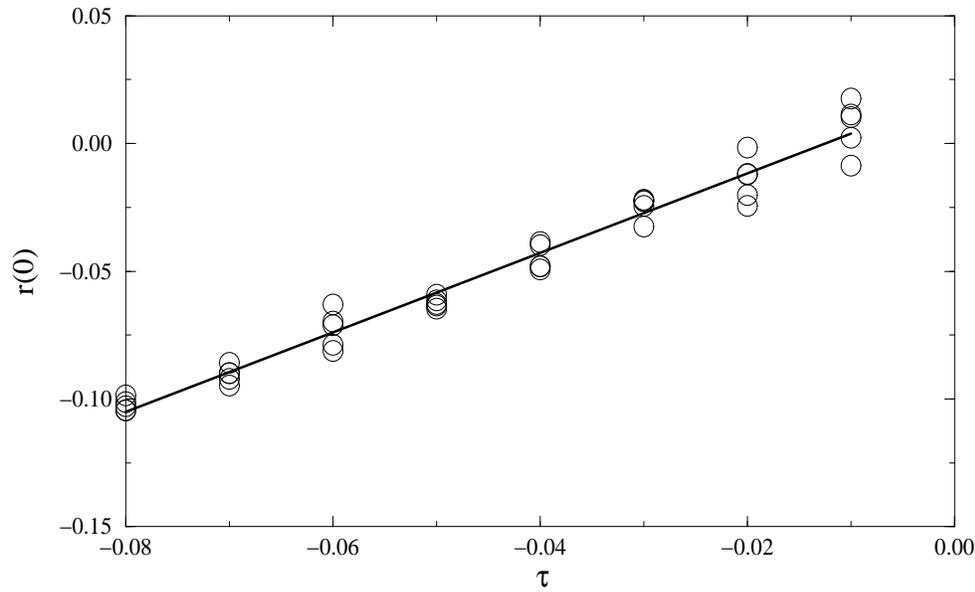}
\vspace{0.3in}
\caption{Growth time as calculated from
fits to a linear theory plotted against the quench depth below the mean
field transition. For short times the system should be described by linear
theory which predicts unstable growth for all quenches below the mean field
transition.
}
\label{Fig:r(0) vs. tau}
\end{figure}

This scenario is confirmed by the simulations. The simulation results
discussed in Section III B confirm that the equilibrium value, $r_{eq}$, is
consistent with the Hartree prediction. Analyzing the early stage dynamics
can verify the value of $r(t)$ at short times. Figure \ref{Fig: S(q0) shorttimes} shows the growth of $S(q_{0})$ as a function of time for various
values of the bare control parameter, $\tau $. To average over noise, five
independent runs where taken for each value of the control parameter. If $
r(t)$ is {\it time-independent}, the early-stage evolution (linear theory)
is described by \cite{CHC}
\begin{equation}
S(q,t)=C_{q}(t,t)=S(q,0)\exp \{-D(q)t\}+{\frac{1}{D(q)}}(1-\exp \{-D(q)t\})
\label{Eq:Linear Theory}
\end{equation}
where $D(q)$ is $D(q)=r_{0}+(q^{2}-q_{0}^{2})^{2}$ and $r_{0}$ is the
time-independent value of $r$. At the peak of the structure factor, $q=q_{0}$
, and $D(q_{0})$ is just $r_{0}$, the value of which can be estimated by
fitting the simulation results shown in Fig. \ref{Fig: S(q0) shorttimes} to 
\ref{Eq:Linear Theory}. The results of such fits are presented in Fig. \ref
{Fig:r(0) vs. tau}. This figure shows that $r_{0}$ and $\tau $ are linearly
related with a slope which is nearly one. This implies that for short times,
the dynamically renormalized parameter is linearly related to the bare
parameter and is not renormalized to positive values for negative values of
the bare parameter. The very early-stage dynamics, therefore, is
characteristic of a system exhibiting unstable growth.

\begin{figure}[h]
\epsfxsize=5.1in \epsfysize=3.2in
\epsfbox{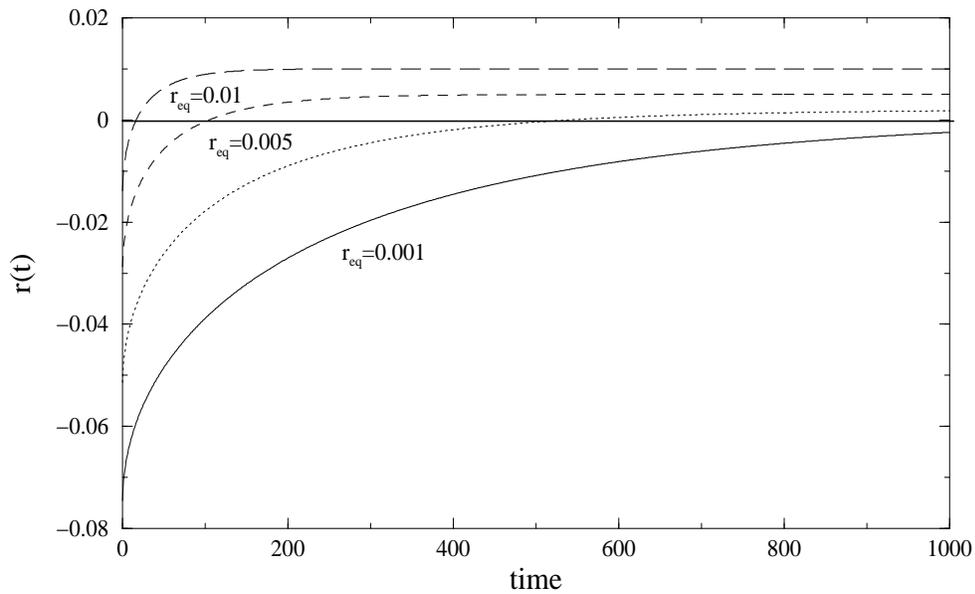}
\vspace{0.3in}
\caption{The time evolution of the mass 
parameter as calculated by dynamic renormalization. $r(t)$ eventually
becomes positive for all quenches below the mean field transition but the
time for that cross over increases dramatically for larger quenches. At
large times $r(t)$ approaches $r_{eq}$.
}
\label{Fig: r(t) vs. time}
\end{figure}

These fits must be considered with some care. In the case where $r$ is not
time dependent, the linear theory describes the system only for short times,
that is times on the order of the natural time scale, $r^{-1}$ [See for
example\cite{GKL94}. For the fastest growing mode, when the data is fit to
different time scales the results obtained are consistent for all time
scales up to $r^{-1}$. For the S-H equation, however, the results obtained are
dependent on the time range. In particular, as the range of time increases
the values obtained for $r$ become less negative. This is consistent with $
r(t)$ growing as a function of time and the results of the fits merely
represent some average value of $r(t)$ for the time range involved. If the
system has been quenched to a negative value of $\tau $, Eq. \ref{r(t)}
indicates that the early-time evolution will exhibit unstable growth. As
time evolves the second term in Eq. (\ref{r(t)}) decreases and the value of $r(t)$
approaches $r_{eq}$ which is always positive. Though the integral in Eq. \ref
{r(t)} is hard to evaluate analytically, numerics can provide some insight
into its behavior. Figure \ref{Fig: r(t) vs. time} shows a numerical
evaluation of the time evolution of $r(t)$ for various values of $r_{eq}$
and $\lambda =0.01$. As confirmed in the statics section above, $r(t)$ will
eventually become positive no matter how deep the quench is, though the time
for this to happen may become very long. As long as $r(t)$ is negative
the system is expected to undergo unstable growth though, since the mass
parameter is time-dependent, the time evolution will be more complicated
than the usual spinodal decomposition scenario. We can define a crossover
time, $t_{cross}$ at which $r(t)$ changes sign, and an average growth time, $
t_{ave}$. The average growth time is just the inverse of $r_{ave}$ defined as

\begin{equation}
t_{ave}^{-1}=r_{ave}=\frac{1}{t_{cross}}\int_{0}^{t_{cross}}r(t^{\prime
})dt^{\prime }  \label{Eq: tave}
\end{equation}

\begin{figure}[h]
\epsfxsize=5in \epsfysize=3.2in
\epsfbox{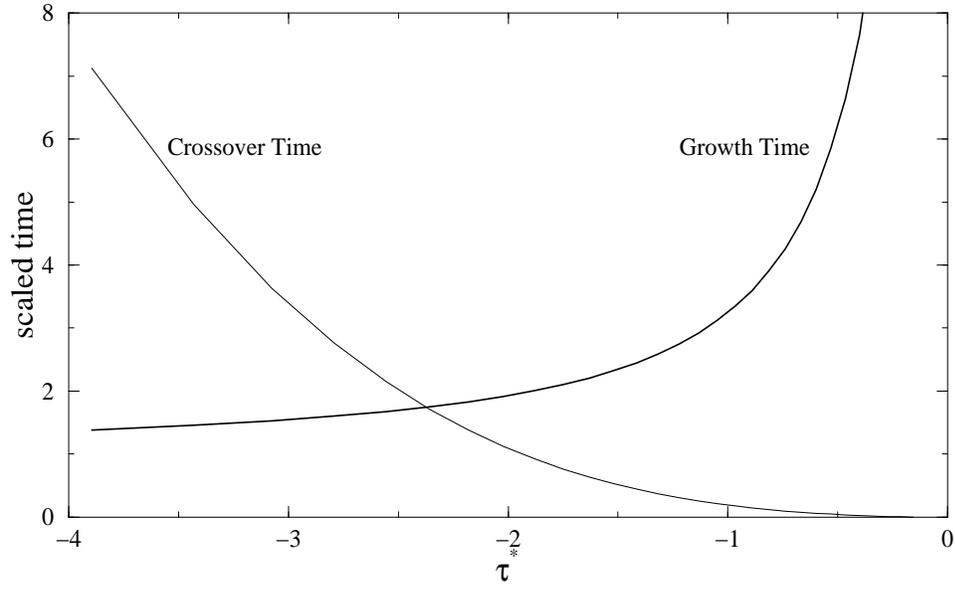}
\vspace{0.3in}
\caption{The calculated crossover time
plotted against quench depth. Also plotted is a characteristic growth time
for the lamellar structure. These results have been scaled using equation
\ref{r Scaling} so that they are $\lambda$-independent. When this time becomes much
less then the crossover time then the lamella has enough time to grow.
Notice that this occurs just above or at the transition temperature, ${\tau}_{trans}$.
}
\label{Fig: tcross vs tau}
\end{figure}

In Fig. \ref{Fig: tcross vs tau}, we show plots of $t_{cross}$ and $t_{ave}$
as a function of the bare control parameter. For shallow quenches $t_{cross}$
is small compared to $t_{ave}$ and the system reaches a well-defined
metastable equilibrium state which is disordered. If $t_{cross}$ becomes
larger then $t_{ave}$, metastability becomes difficult to define. The system
takes longer to equilibrate in the metastable disordered phase than it takes
to grow lamellar structures. We can use the condition $t_{cross} = t_{ave}$
to define a crossover temperature ${\tau}_{dyn}^*$. Above this temperature
the disordered phase quickly becomes locally stable and a nucleation event
is needed for the formation of the lamellar structures. Below ${\tau}
_{dyn}^* $, the system is expected to evolve continuously towards the
lamellar phase without any evidence of metastability of the disordered
phase. Fig. \ref{Fig: tcross vs tau} is in sharp contrast to standard
Ginzburg-Landau (G-L) theory.  For the time dependent G-L equation, the presence 
of noise suppresses the critical point.
In the region where the bare parameter is negative but the
renormalized parameter is positive, $t_{cross}$ is always smaller then
$t_{ave}$ and so there is no unstable growth until the true
critical point is reached.  

For the system under consideration here, different scenarios are
possible depending on the relationship of this dynamical crossover
temperature to the first-order transition temperature,
${\tau}_{trans}^*$ , at which the free energy of the lamellar phase
becomes lower that that of the disordered phase. If ${\tau}_{trans}^*$
is higher than ${\tau}_{dyn}^*$, there will be a regime of
temperatures over which the the system will undergo nucleation and
growth. On the other hand if ${\tau}_{trans}^*$ is lower than
${\tau}_{dyn}^*$, then there is no nucleating regime and one would
observe only unstable growth, albeit of an unusual nature since the
free-energy surface is evolving with time. From figure \ref {Fig:
tcross vs tau} we can see that $t_{cross}$ is small only for very
shallow quenches below the mean-field transition. The results of our simulations,
presented in the next section, are in qualitative agreement with this
scenario. We would like to emphasize that the growth dynamics
described above are qualitatively different from rapid nucleation in
which the system quickly reaches the stable equilibrium state. A large
value of $t_{cross}$, on the other hand, indicates a type of
ergodicity breaking as the system takes a very long time to reach
equilibrium.

All of our numerical results indicate that ${\tau}_{trans}^*$ predicted from
Hartree is lower than ${\tau}_{dyn}^*$ and they are remarkably close to each
other. We have been unable to come up with a simple relationship between
these two temperatures and, therefore, can only interpret the similarity of
the two as a remarkable coincidence. The dynamical crossover is deduced from
time scales which characterize the evolution of the free-energy surface
while the transition temperature is deduced from a comparison of the depth
of the two wells. It is not clear why the two temperatures should be similar
in magnitude. It can be argued that there is only a single parameter, $
\lambda$, controlling the scale of fluctuations and, therefore, the two
temperatures should be related, however, there is no obvious argument to
suggest that they should be identical.

The picture emerging from the dynamical renormalization is a natural
extension of the effect of fluctuations on the static results of the
Brazovskii model. Following a quench from a relatively high temperature to a
temperature where $\tau $ is negative, the system is in a locally unstable
region (top of a hill). As the fluctuations grow with time, the non-linear
terms characterized by $\lambda$ become important and they renormalize the
curvature of the hill. This scenario is quite different from the usual
picture of evolution in an adiabatic potential.

\subsection{Simulation Results for Late Times}

Further evidence for the dynamical scenario presented in the previous
section comes from examining the long time evolution of the peak of
the structure factor obtained from our numerical simulations and
correlating that to snap shots of the system as it evolves. Figure
\ref{Fig: S(q0) long time} is a plot of the amplitude of the structure
factor peak as it evolves.  For shallow quenches, $\tau >-0.04, $the
peak grows to an equilibrium value and does not evolve further. These
values are consistent with the equilibrium values reported above in
the section on statics. For deeper quenches the peak amplitude grows
quickly and then appears to saturate; however careful examination
shows that the value continues to grow very slowly. This is consistent
with late stage domain growth which as been studied extensively by
Elder et. al.  for 2-D systems \cite{EGV1995}. The last feature in the
system is a final rise to an equilibrium value. This rise is a finite
size affect is caused by the majority domain finally taking over the
entire system. This being the case, the peak would not be expected to
grow much since the difference between the value before and after this
final evolution should only represent the surface area between the
different domains.

\begin{figure}[h]
\epsfxsize=2.55in \epsfysize=2.06in
\epsfbox{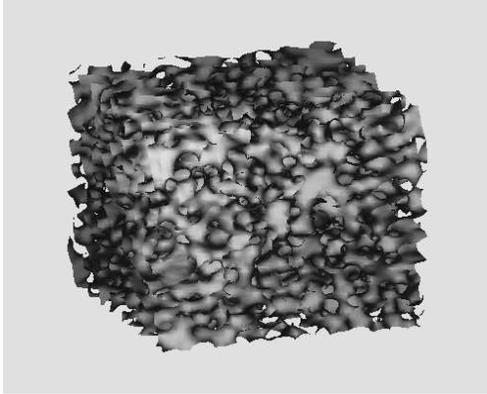}
\caption{Time = 1000}
\label{Fig:System snapshots1}
\end{figure}

\begin{figure}[h]
\epsfxsize=2.55in \epsfysize=2.06in
\epsfbox{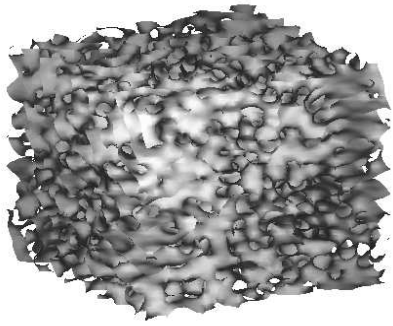}
\caption{Time = 2000}
\label{Fig:System snapshots2}
\end{figure}

\begin{figure}[hy]
\epsfxsize=2.55in \epsfysize=2.06in
\epsfbox{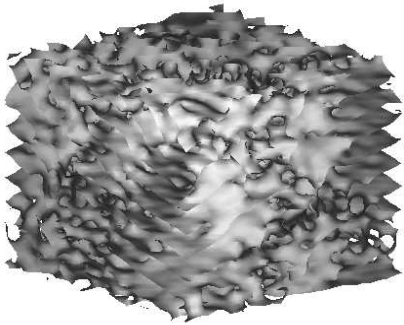}
\caption{Time = 3000}
\label{Fig:System snapshots3}
\end{figure}

\begin{figure}[h]
\epsfxsize=2.55in \epsfysize=2.06in
\epsfbox{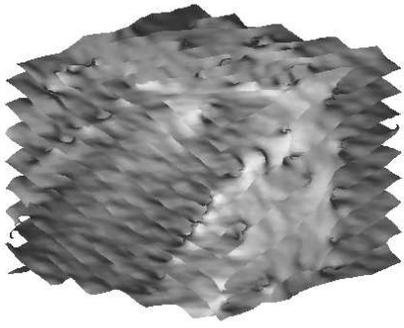}
\caption{Time = 4000}
\label{Fig:System snapshots4}
\end{figure}

\begin{figure}[h]
\epsfxsize=2.55in \epsfysize=2.06in
\epsfbox{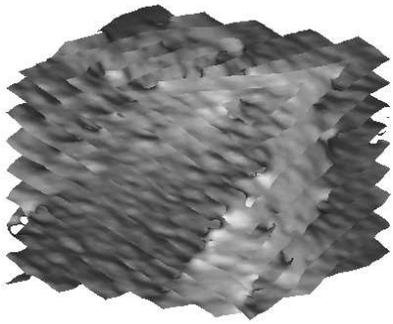}
\caption{Time = 5000}
\label{Fig:System snapshots5}
\end{figure}

For a quench depth of $\tau =-0.07$ this time evolution can be compared to a
series of system snapshots shown in figures \ref{Fig:System snapshots1}-
\ref{Fig:System snapshots5}. These
figures represent $\phi =0$ isosurfaces which would be the boundaries
between the different microphases of the system. For early times the system
appears to be very disordered while at later times domains of ordered
lamellar structures begin to appear. Just before the final convulsive
growth, at a time $t=4000$, there still appear to be a few different ordered
domains while just after the final growth, at $t=5000$, only one domain
appears to be left in the system. Thus we still have a consistent picture of
slow but continuous domain growth in the system which is governed by the
evolution of the 2nd order term as it goes from negative values to positive
values.

\section{Future Work}

\begin{figure}[h]
\epsfxsize=5.1in \epsfysize=3.2in
\epsfbox{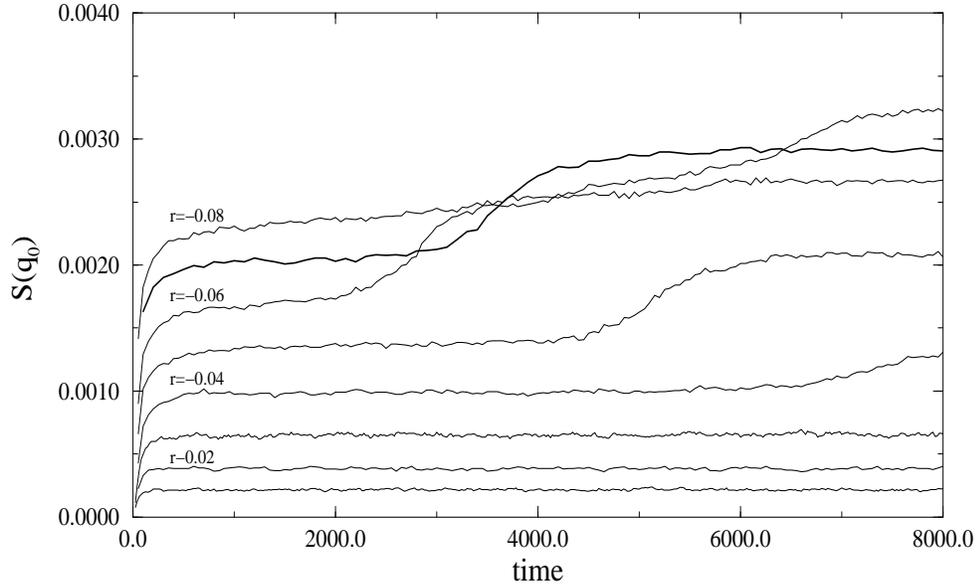}
\vspace{0.3in}
\caption{The evolution of the structure
factor peak as a function of time for different quench depths. For shallow
quenches above the real transition temperature $S(q_{0})$ grows quickly to
its equilibrium value. For quenches to below the transition temperature the
system grows quickly at early times and then reaches a late stage regime
where it evolves by domain growth and $S(q_{0})$ grows as a small power law
in time. The highlighted line is a quench to $r=-0.07$. Snapshots of this
system are presented in the next figure.
}
\label{Fig: S(q0) long time}
\end{figure}

In the work presented above we have shown that analysis of the static
free energy does not always provide an adequate description of the
fluctuation driven dynamics. Though figure \ref{Fig: S(q0) long time}
shows features which are reminiscent of a first order phase
transition, dynamical analysis of the renormalized coefficients
suggest that a more complicated evolution is taking place. A similar situation
occurs for the superconducting transition which is also a fluctuation-driven
first-order transition\cite{Gold91}. 

In the present 
work we have paid close attention to the time evolution of the mass
parameter and its effect on the evolution of the system at early
times. The dynamical renormalization results also become tractable at
times large compared to $t_{cross}$ when the system has reached
equilibrium. As pointed out earlier, one enters this regime fairly
quickly for shallow quenches. At this stage, the higher order terms in
$\Gamma _{dyn}$ become important. The most interesting aspect of the
renormalized theory is that the fourth-order term becomes non-local in
time \cite{ICG99}:
\[
\Gamma _{dyn}(4)\simeq \int d1d2 \Phi ^{2}(1)K(1-2)\Phi ^{2}(2) 
\]
and the integral of the kernel over all times leads to the Brazovskii result
for the renormalized fourth-order term. One can derive an ``effective''
Langevin equation for the fields $\phi $ from the renormalized $\Gamma
_{dyn} $, and the above equation implies that this Langevin equation has
``memory'' effects which appear in the non-linear term. This would lead to
unusual behavior of the equilibrium correlations functions and might provide
a distinctive signature of fluctuation-driven first-order transitions. We
are in the process of investigating the effective equation for $\phi$
numerically. One obvious consequence of the non-local term is that the
barrier to nucleation is dynamic and it is not valid to think of nucleation
as tunneling through a fixed barrier. The situation is closer to the problem
of quantum tunneling for many degrees of freedom where one also finds an
effective equation which is non-local in time\cite{Hedegard}.

Another area of interest is the role of defects  in the phase
transition of the system. These will be important for very shallow
quenches to just below the transition point and we have noted there
presence when looking at snapshots of the system. These defects are
reminiscent of the perforated lamellar phase described in reference
\cite{Bib:Blockcopolymer}. One way to study this is to use a
directional order parameter measure introduced by Bray et al.\cite
{bib:Bray98} to study the 2D problem. This order parameter is similar
to the order parameter for complex fluids and provides directional as
well as density information. What it may show is that the local
gradient terms are enhanced for negative values $\tau $ suggesting
that perhaps there is a disordered to nematic transition in this
region which others have suggested for 2D \cite{EGV1995}.

We would like to close with a discussion of possible experimental systems
where this dynamical scenario can be studied experimentally. Since the
effective $\lambda$ is extremely small in Rayleigh-Benard systems the regime
where the disordered state is metastable will probably be unaccessible. In
di-block copolymers, however, there should be a significant temperature
range, below the mean-field transition, where the disordered state is locally
stable and a study of the equilibrium, two-time correlation function should
reveal the existence of the non-linear memory term. It might also be
possible to observe unusual nucleation if the system parameters allow for $
\tau_{trans}^*$ to be higher than $\tau_{dyn}^*$.

\vskip 1in

\pagebreak

\pagebreak

\vspace{0.3in}

\pagebreak

\pagebreak
\end{document}